\documentclass[sigconf]{acmart}
\usepackage{graphicx} 
\usepackage{algorithmic}
\usepackage{textcomp}
\usepackage{url}
\usepackage{enumitem}
\usepackage{xspace}
\usepackage{soul}
\usepackage{stfloats}
\usepackage{xurl}
\usepackage{multirow}

\usepackage{tikz}

\copyrightyear{2026}
\acmYear{2026}
\setcopyright{cc}
\setcctype{by}
\acmConference[ICSE-Companion '26]{2026 IEEE/ACM 48th International Conference on Software Engineering}{April 12--18, 2026}{Rio de Janeiro, Brazil}
\acmBooktitle{2026 IEEE/ACM 48th International Conference on Software Engineering (ICSE-Companion '26), April 12--18, 2026, Rio de Janeiro, Brazil}
\acmPrice{}
\acmDOI{10.1145/3774748.3787627}
\acmISBN{979-8-4007-2296-7/2026/04}

\title{Multi-Level Testing of Conversational AI Systems}

\author{Elena Masserini}
\email{elena.masserini@unimib.it}
\orcid{0009-0002-6969-1500}
\affiliation{
  \institution{Ph.D. student, year 1 out of expected 3 \\University of Milano-Bicocca, Milan, Italy}
  \country{Supervised by Prof. Daniela Micucci and Prof. Leonardo Mariani}
}

\begin{document}

\begin{abstract}
Conversational AI systems combine AI-based solutions with the flexibility of conversational interfaces. However, most existing testing solutions do not straightforwardly adapt to the characteristics of conversational interaction or to the behavior of AI components.
To address this limitation, this Ph.D. thesis investigates a new family of testing approaches for conversational AI systems, focusing on the validation of their constituent elements at different levels of granularity, 
from the integration between the language and the AI components, to individual conversational agents, up to multi-agent implementations of conversational AI systems. \end{abstract}

\maketitle

\section{Introduction}\label{sec:introduction}

In recent years, conversational AI systems have gained significant popularity, becoming increasingly integrated in everyday life and widely applied in a multitude of domains, like e-commerce, banking, healthcare, and education \cite{chatbot-statistics}. These systems are designed to interact with users through human-like interactions that facilitate information exchange, and ease access to services. 
Conversational AI systems can be implemented as single agents \cite{zhu-ca, zhang-ca} or as cooperating multi-agent systems \cite{gody-mas, wang-mas} that collectively deliver the required functionalities. 

Despite their growing adoption and the satisfaction reported by users and companies~
\cite{forethough-statistics}, conversational AI systems still exhibit numerous issues. Common problems include users having to repeat sentences multiple times \cite{forethough-statistics}, and seek for human assistance \cite{clutch}.
More severe cases include the diffusion of incorrect and misleading information, such as the Air Canada AI assistant lying to passengers~\cite{aircanada-chatbot}, or a NYC business AI assistant suggesting illegal practices~\cite{nycs-chatbot}. In addition, AI-driven assistants have exposed critical security vulnerabilities, such as unauthorized access to voice history in Alexa devices~\cite{alexa-security}.


These shortcomings highlight the need for better quality assurance  strategies that can thoroughly assess the quality of conversational systems. Traditional testing strategies cannot be straightforwardly adapted to conversational AI systems, as they introduce several difficulties, like: user requests and system responses are natural language sentences, that must be interpreted to establish the correctness of the test;  
the same user input (a sentence) can be expressed in potentially unlimited ways, complicating comprehensive test coverage; 
the test oracle must accept different semantically equivalent correct responses; 
system assessment must consider not only the responses provided, but also the actions performed (e.g., on external services or devices) and the internal state changes; the non-deterministic nature of these systems requires probabilistic evaluation methods rather than traditional binary approaches.

Over the last years, some approaches addressed quality assessment of conversational AI systems, targeting conversations, functionalities, and the interactions between agents. 
%
%
Botium~\cite{botium} is a popular multi-platform tool that generates and executes simple conversational tests derived from the training phrases of the tested system. Subsequent research has extended Botium to exercise slightly more complex conversational scenarios~\cite{canizares2024coverage, gianni2025test}. 
Other approaches, such as Charm \cite{bravo2020testing} and BoTest \cite{8718238}, investigated the generation of robustness tests by using, for example, synonyms and paraphrases. 
Conversational agents can also be implemented by connecting LLMs with external APIs~\cite{basu2024apiblend}. In this context, Arcadinho et al. explored the idea of generating conversations whose goal is triggering internal API calls~\cite{arcadinho2024}. 
Lastly, as multi-agent architecture is becoming a popular paradigm \cite{agentic-ai-market}, recent works specifically target multi-agent conversational AI systems to test their reliability under challenging conditions, injecting faults within the single agents \cite{pmlr-v267-huang25ay} or in their operating environment \cite{chaos-eng}. 

However, the ability to systematically explore the huge, virtually unlimited, conversational space, thoroughly exercise functionalities, and scale to multi-agent systems remains limited.



To advance research in this area and address these open challenges, this Ph.D. thesis aims to define automated testing strategies that support the creation of trustworthy and reliable conversational AI systems. 
The research plan is articulated across three complementary levels of abstractions, moving from the interactions of individual components to the behavior of the system as a whole. The first level, \textit{service-interaction testing}, concentrates on exercising interactions between the language component (e.g., an LLM or an NLP pipeline) and the services it relies on.
The second level, \textit{agent testing,} examines whether a conversational agent behaves correctly when engaging with users or with other agents. The third level, \textit{multi-agent system testing}, focuses on validating the overall behavior of the conversational AI system and assessing whether it fulfills its intended requirements.


\section{Conversational AI Systems Architecture}\label{sec:architecure}

\begin{figure}[H]
    \vspace{-2mm}
    \centering
\includegraphics[width=\columnwidth]{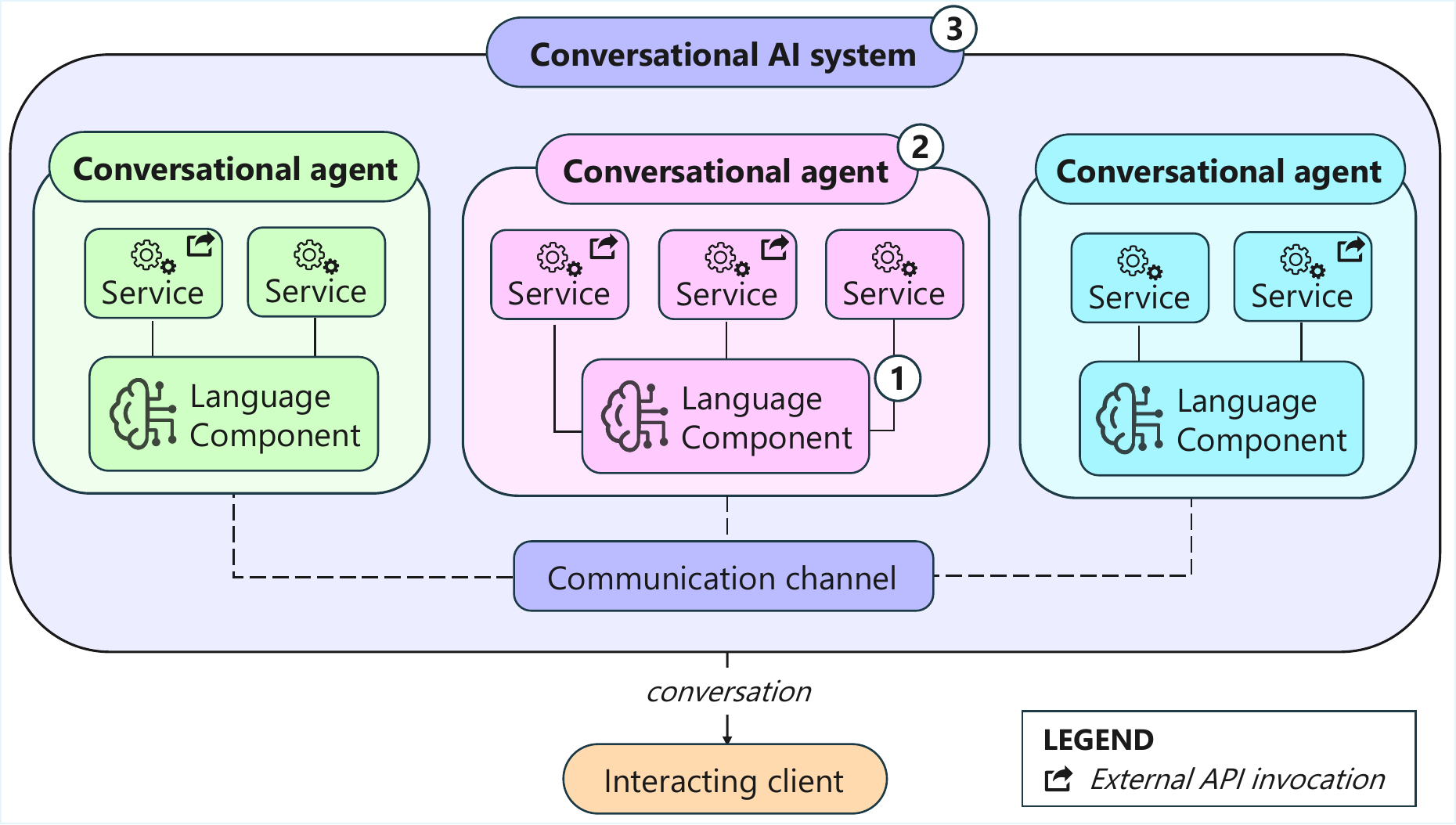}
    \caption{Multi-agent conversational AI system architecture.}
    \label{fig:mas-architecture}
    \vspace{-2mm}
    \Description{Diagram showing the architecture of a multi-agent conversational AI system made of three agents. Each agent contains a language component and a set of services that may invoke external APIs. Connections between the language component and each service are labeled as 1, while each agent is labeled as 2. The whole conversational AI system is labeled as 3 and interacts with an interacting client through conversation.}
\end{figure}

As shown in Figure ~\ref{fig:mas-architecture}, conversational agents typically consist of two main elements: a \textit{language component}, and a set of \textit{services}. 

The \textit{language component} handles the conversational aspects of the interaction, such as processing and interpreting requests, extracting input parameters, and generating responses. As discussed in~\cite{SINGH2025100128}, language components 
can be \textit{rule-based}, relying on pattern-matching rules; \textit{retrieval-based}, supported by machine learning models that retrieve responses from a predefined set; or \textit{generative-based}, implemented through deep learning models and LLMs that generate responses dynamically. 

A \textit{service} exposes the operations that the language component can invoke to carry out task-oriented actions, defining the functionality that can be delivered by the agent.
For instance, a NLP pipeline or a LLM may invoke a REST API to create an event in a shared calendar. Services can be implemented internally within the agent or provided externally as third-party APIs.


Conversational AI systems may consist of a single conversational agent or a collaborative network of multiple agents. In the latter case, the system operates as a \textit{multi-agent conversational system}, where agents exchange information, coordinate their behaviour, and distribute responsibilities to jointly address the interacting client’s (i.e., a human user or an external system) goals.



\section{Research Approach}\label{sec:approach}

This Ph.D. thesis aims to study methods that can thoroughly validate conversational AI systems, considering the novel and prominent role of language processing components, services, and agents, which may interact in unexpected ways. To address this problem, this work targets different granularity levels that represent increasing levels of complexity of the interactions, with a progressively larger number of elements involved.

Referring to the architecture of conversational AI systems shown in Figure~\ref{fig:mas-architecture}, the research will address: \textit{service-interaction testing}, whose objective is to assess the interaction between language components and services (see \{1\} in Figure~\ref{fig:mas-architecture}); \textit{agent testing}, whose objective is to assess whole agents (see \{2\} in Figure~\ref{fig:mas-architecture}); and \textit{multi-agent system testing}, whose objective is to assess the integration of multiple agents (see \{3\} in Figure~\ref{fig:mas-architecture}). 
Note that fine-grained aspects below the service-interaction level (e.g., internal elements of the language component and the service implementation) are outside the scope of this work, as they could be addressed with state-of-the-art approaches.

\smallskip \noindent \emph{Service-interaction testing}
aims to thoroughly exercise the interactions between the language component and the services, to ensure that the system can deliver the intended functionality. The language component must interpret user sentences and convert them into the invocation of the appropriate services, selecting the correct operations in the right sequence and providing them with the proper values. 
To systematically test this integration, 
we plan to model test generation as a search problem, where proper input sentences must be searched to trigger enough interactions between language components and services. In particular, we will study feedback-directed gray-box strategies that guide input sentences generation based on the service invocations triggered by the produced inputs. These gray box strategies rely on the observation of interactions (e.g., API coverage, parameter coverage), which is generally feasible, without assuming access to the internal details of the tested components, which might sometimes be infeasible. Since language plays a relevant role, we plan to integrate LLM-based algorithms in the search process to enable a deeper and more targeted exploration, to obtain diverse and semantically meaningful conversational tests. 


\smallskip \noindent
\emph{Agent testing} 
aims to validate whether a conversational agent fulfills the intended goals from the interacting client’s perspective. 
The challenge is that conversational AI systems are usually not available with well-formalized requirements, especially linking conversations with functionalities. We will address this challenge in two ways. On the one hand, we will study how to enrich requirement specifications with information about conversations that can be used to derive more meaningful test cases. On the other hand, we will investigate the use of metamorphic testing approaches~\cite{cho2025metamorphictestinglargelanguage}, which are particularly suitable when only sparse or no knowledge about the expected behavior is available~\cite{bozic-metamorphic}. Metamorphic relations will allow the definition of relationships between input transformations and corresponding output behavior, supporting the automatic derivation of test cases.

\smallskip \noindent
\emph{Multi-agent system testing} 
aims at assessing the correct behavior of a conversational AI system, considered as a whole. To exercise the entire system, testing approaches should design and execute complex scenarios that involve interactions between multiple agents, and observe how they coordinate their actions to achieve shared goals. A central challenge when testing a scenario that involves multiple agents is ensuring that each agent performs the right actions at the right time to produce the expected outcome. To address this challenge, we will investigate test generation strategies based on both planning \cite{planning} and orchestration \cite{orchestration}. AI planning can be used to generate test workflows based on the goal of the multi-agent conversational system. To generate the concrete tests implementing the workflow, orchestration techniques can be exploited, jointly with the injection of special testing agents (e.g., testing agents and mocking agents) into the system to exercise specific behaviors, including erroneous and rare situations.

\section{Work Plan}\label{sec:work-plan}




The doctoral research spans three years and is organized in three main stages, approximately lasting one year each, tackling progressively more complex aspects of conversational AI systems.
In the first stage, we will study test case generation techniques that target the integration between the language component and the services. In the second stage, we will investigate test case generation for conversational agents, focusing on techniques capable of revealing issues emerging from interactions among the agent's components. In the third stage, we will address test case generation for multi-agent conversational systems, targeting problems that arise from the integration of multiple conversational AI agents.

In the initial phase of our research, we built a dataset of RASA\footnote{https://rasa.com/} 
and Dialogflow\footnote{https://docs.cloud.google.com/dialogflow/docs}
conversational agents~\cite{masserini2025brasato}, which will be used as experimental subjects in the evaluation of the proposed techniques. We also conducted a preliminary study on the effectiveness of test cases generated by Botium, which highlighted some limitations of the tool. Moreover, we worked on defining a mutation testing approach for conversational systems~\cite{clerissi2025towards, dialogflow-repo}, so that test case generation techniques can be assessed using fault-based metrics, which are valuable indicators of the quality of a test suite. Lastly, we are currently working on automatic test case generation for retrieval-based conversational agents using LLMs and on the definition of a LLM-driven oracle to evaluate the correctness of a given interaction based on system specifications and sample conversations. 

The evaluation plan involves expanding the initial dataset of conversational AI systems to include also open-source generative agents and multi-agent systems. The designed test case generation techniques will be compared to baseline methods, such as Botium and Charm, using quality metrics based on conversational and code coverage and fault metrics such as mutation score and real faults revealed. Beyond outperforming competing approaches, the goal is to design techniques that can reveal faults that matter in practice (e.g., faults that developers are willing to address if reported) and that practitioners are interested to use.

\section{Expected Contributions}\label{sec:contributions}

This thesis is expected to deliver five main contributions: 
(1) A curated dataset of conversations AI systems, both consisting of individual agents and multi-agent systems, that can be used to advance research in quality assurance of conversational AI systems; (2) Feedback-directed testing methods to thoroughly validate how language components interact with services, to provide the intended functionality;
(3) Specification-driven testing methods to validate individual conversational agents against requirements and metamorphic relations; (4) Testing and mocking agents that can be injected into multi-agent conversational systems to validate interactions and collaborations among agents; (5) Empirical evidence that the proposed methods can be used to enhance the quality of conversational AI systems, timely revealing faults occurring at different granularity levels, from the integration to the system level.

\bibliographystyle{ACM-Reference-Format}
\bibliography{references}

\end{document}